\documentclass[journal,twocolumn]{IEEEtran}
\usepackage{cite,graphicx,amssymb,amsmath,color,textcomp}
\usepackage{extarrows,multirow,multicol}
\usepackage{bm} %\bm{}
\usepackage{algorithm }
\usepackage{algorithmic }

\newtheorem{remark}{\underline{Remark}}%[section]

\newlength{\figwidth}
\begin{document}
\title{
\huge Channel Estimation for Ambient Backscatter Communication Systems with Massive-Antenna Reader}
\author{
\vspace{-0mm}
Wenjing Zhao,  Gongpu Wang, Saman Atapattu, {\it Senior Member, IEEE},\\
Ruisi He, {\it Senior Member, IEEE}, and Ying-Chang Liang, {\it Fellow, IEEE}
\thanks{Copyright (c) 2015 IEEE. Personal use of this material is permitted. However, permission to use this material for any other purposes must be obtained from the IEEE by sending a request to pubs-permissions@ieee.org.}
\thanks{This work is supported in part by  the Fundamental Research Funds
for the Central Universities under Grant 2018YJS047 and 2016JBZ006, in part by
the Australian Research Council (ARC) through the Discovery Early Career
Researcher (DECRA) Award DE160100020, in part by Key Laboratory of Universal Wireless Communications (BUPT), Ministry of Education, P.R.China under Grant KFKT-2018104, and in part by the National Natural Science Foundation of China under grant 2016YFE0200900, 61571037 and 61871026.
(Corresponding author: Gongpu Wang)}
\thanks{W. Zhao, G. Wang and R. He are with Beijing Key Lab of Transportation Data Analysis
and Mining, School of Computer and Information Technology, Beijing
Jiaotong University, China (e-mail: \{wenjingzhao,~gpwang,~ruisi.he\}@bjtu.edu.cn).

S.~Atapattu is with the Department of Electrical and Electronic Engineering,  University of Melbourne, Parkville, VIC 3010, Australia (e-mail: saman.atapattu@unimelb.edu.au).

Y.-C. Liang is with the Center for Intelligent Networking and
Communications, University of Electronic Science and Technology of China,
Chengdu, China (e-mail: liangyc@ieee.org).
}
\vspace{-7mm}
}

\maketitle
\begin{abstract}
Ambient backscatter, an emerging green communication technology,  has aroused great interest from both academia and industry.
One open problem for ambient backscatter communication (AmBC) systems is channel estimation for  a massive-antenna reader.
In this paper, we focus on channel estimation problem in AmBC systems with uniform linear array (ULA) at the reader which consists of large number of antennas.
We first design a two-step method to jointly estimate channel gains and direction of arrivals (DoAs), and then  refine the   estimates through angular rotation.
Additionally, Cram\'{e}r-Rao lower bounds (CRLBs) are derived for both the modulus of the channel gain and the DoA estimates.
Simulations  are then provided to validate the analysis, and to show the efficiency of the proposed approach.
\end{abstract}


\vspace{-4mm}
\section{Introduction} \label{S1}
Aiming to enable easy access and interaction among numerous computing devices such as sensors and  Internet of Things (IoT) will play a vital role in the future communication paradigm \cite{zhou2014mwc, Fuqaha2015jst}.
Recently,  there have been a variety of research directions on IoT.
 Among them, how  sustainable
 and reliable energy can be supplied to  large-scale deployments of IoT devices is an interesting and challenging problem nowadays. Since ambient backscatter leverages environmental  radio frequency (RF) signals to enable battery-free devices to communicate with each other, it has high potential to offer a solution for the energy problem in IoT systems \cite{liu2013sigcomm}.
An ambient backscatter communication (AmBC) system typically consists of a RF source, reader and tag. Before modulating its own binary data,  the tag first harvests energy from RF signals, which thus exempts the tag from energy constraint.
 Then,  the tag loads  bit~`1'  by reflecting the incident RF signals and    bit~`0' by absorbing them. By using certain detector, such as maximum-likelihood (ML) detector,  the reader demodulates the bit information  accordingly  \cite{tao2019wcom}.

 The majority of  existing theoretical studies on AmBC,  related to signal detection \cite{tao2018wcom, tao2019wcom}
  and the references therein, performance analysis \cite{zhao2018coml,li2018tvt}  and multiple access scheme \cite{liu2018wcom},  assume perfect channel state information (CSI).
 In reality, precise knowledge of full CSI is  not always available, especially with strict energy constrained IoT systems.
  Other than contributing to signal detection, perfect CSI plays a key role in  transceiver design and security improvement \cite{zou2016network, zhu2016tvt, zhu2016access}.

Traditionally, instantaneous channel coefficients can be obtained through the channel estimation and CSI exchange procedure in every coherence time. However, accurate channel estimation  may only be possible with reasonably long enough training signals, which costs significant time and power, especially with strict energy-constrained massive-antenna IoT systems. Since the tag in an AmBC system  merely modulates its signals by reflecting  the incident signals, it is unable to transmit additional training or pilot signals. Therefore, traditional channel estimation techniques may not be directly applied to AmBC systems. Moreover, while RF signals are usually unknown to the both reader and tag, the inconsistencies of channels at reflective and absorptive states also pose great challenges on channel estimation.
Taking these into account, an expectation maximization (EM) based estimator is designed to acquire the modulus values of  channels in an AmBC system with a single-antenna reader in  \cite{ma2018coml}.
In  a multiple-antenna reader circumstance,  an  approach on the strength of eigenvalue decomposition (EVD) \cite{zhao2018iccc} is adopted to retrieve channel parameters. Nevertheless,  the complexity of EVD is prohibitive in AmBC systems with massive-antenna \cite{ma2018wcom, ma2019com, Atapattu2019twc, wang2018tsp} reader, which motivates our work.
In this paper, we tackle the channel estimation problem in AmBC systems with massive-antenna reader  having an uniform linear array (ULA).
Together with least-square (LS) method,
an estimator resorting to discrete Fourier transformation (DFT) \cite{fan2017jsac} and angular rotation operation is presented to  collectively figure out the direction of arrivals (DoAs) and channel gains. To the best of our knowledge, this is the first work which considers on  channel estimation for an AmBC system with massive antennas.

\emph{\textbf{Notations}}: We use   boldfaced lowercase for vectors and boldface uppercase for matrices.
  The  transpose and the  inverse of matrix $\textbf{X}$ are  denoted by $\textbf{X}^T$  and $\textbf{X}^{-1}$, respectively.
  $[\textbf{X}]_{ij}$ indicates the $(i,j)$th element of matrix $\textbf{X}$, and $\mathbf{x}_i$ indicates the $i$th element of vector $\textbf{x}$.
  $\text{diag}\{\mathbf{x}\}$ denotes a square diagonal matrix with the elements of $\mathbf{x}$ on the main diagonal. The identity matrix is denoted by $\textbf{I}$.
 $\mathbf{x} \sim \mathcal{CN}(\bm{\mu}, \bm{\Sigma})$ denotes that $\mathbf{x}$ is a circularly symmetric complex
Gaussian (CSCG) vector with mean $\bm{\mu}$ and covariance matrix
$\bm{\Sigma}$.
$\|\mathbf{x}\|$ represents the $2-$norm of vector
$\mathbf{x}$. $\lfloor x \rceil$ rounds $x$ to the nearest integer, and $\mathbb{E} \{x\}$ means the statistical expectation of $x$.
$\mathfrak{Re}\{x\}$ and $\mathfrak{Im}\{x\}$  denote the real part and the imaginary part of $x$, respectively.
\vspace{-2mm}
\section{System Model} \label{S2}

\begin{figure}
\vspace{-2mm}
  \centering
  \includegraphics[height=48mm,width=75mm]{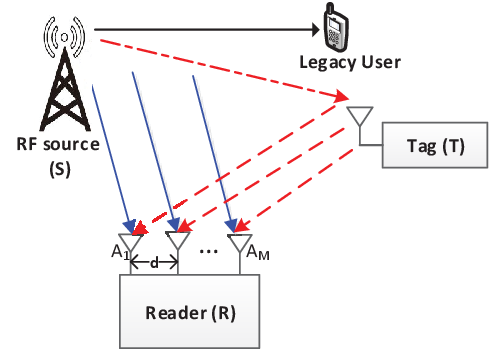}\\
  \caption{An ambient backscatter communication (AmBC) system with a massive-antenna reader.}\label{SystemModel}
  \vspace{-2mm}
\end{figure}
\vspace{-1mm}

As shown in Fig.~\ref{SystemModel}, we consider an AmBC system  with a RF source ($S$), a   reader ($R$) equipped with $M$ antennas in the form of ULA,
and a passive  tag ($T$) with single antenna. %as illustrated in Fig.~\ref{SystemModel}.
The reader not only receives signal from the RF source directly,
but also collects signal backscattered from the tag.
The  tag  first harvests energy  from the RF  signals. By intentionally changing its  load impedance, the tag then piggybacks its information bits over ambient RF carriers to backscatter outside  or to absorb  inside the received signals.

Let $s(n)$ be the signals from the RF source with power $P_s$ and $B(n) \in \{0,1\}$ the modulated signal  at the tag which keeps unchanged during $N$ consecutive RF signals.
Define $\theta_0 \in [-\frac{\pi}{2}, \frac{\pi}{2}]$ and $\theta_1 \in [-\frac{\pi}{2}, \frac{\pi}{2}]$ as the signal azimuth angles or DoAs of paths $S-R$ and $T-R$, respectively.
Denote channel gains of $S-R$, $S-T$ and $T-R$ as ${h}_{sr}$, $h_{st}$ and ${h}_{tr}$, respectively.\footnote{The channel $h_{sr}$ is the traditional source to reader channel.}
The attenuation factor inside the tag is denoted as $\eta \in (0,1]$.
Then, the received signal at the reader is \cite{wang2016tcom}
\begin{align}
\mathbf{y}(n)&= \mathbf{h}_{sr}s(n) + \mathbf{h}_{tr} \eta h_{st} s(n)B(n)+\mathbf{w}(n) \notag\\
&=\mathbf{h}s(n)+\mathbf{w}(n) \label{Rx}, %\label{EqualRx2}
\end{align}
where the equivalent AmBC channel is
\[\mathbf{h}=\mathbf{h}_{sr}+\eta h_{st} \mathbf{h}_{tr} B(n),\]
$\mathbf{h}_{sr}=h_{sr}[1,e^{j \frac{2 \pi d}{\lambda}\sin \theta_0},\cdots,e^{j \frac{2 \pi d}{\lambda}(M-1) \sin \theta_0}]^T$,
$\mathbf{h}_{tr}=h_{tr}[1,e^{j \frac{2 \pi d}{\lambda}\sin \theta_1},\cdots,e^{j \frac{2 \pi d}{\lambda} (M-1) \sin \theta_1}]^T$,  and $\mathbf{w}(n)$ is CSCG noise vector distributed as $\mathbf{w}(n) \sim \mathcal{CN}(\bm{0}, \sigma^2 \mathbf{I})$. Here, $d$ is the distance between two
adjacent antennas and $\lambda$ is the wave length of the RF signal.
Assume the delay distance at the $m$th antenna is $(m-1)d \sin \theta_i$ for $i\in \{0,1\}$ compared to the first antenna. Then it can be noticed that the equivalent channel  at the $m$th antenna is
 \begin{align}\label{singlechannel}
 h_m &=h_{sr} e^{j \frac{2 \pi d}{\lambda} (m-1) \sin \theta_0} + \eta h_{st}h_{tr}B(n) e^{j \frac{2 \pi d}{\lambda} (m-1) \sin \theta_1}\notag \\
 &=h_0 e^{j \frac{2 \pi d}{\lambda} (m-1) \sin \theta_0} + h_1 B(n) e^{j \frac{2 \pi d}{\lambda} (m-1) \sin \theta_1},
 \end{align}
 where $h_0=h_{sr}$ and $h_1=\eta h_{st}h_{tr}$.
{\remark
Since $h_m$ is a function of the modulated bit at the tag and channels $h_{sr}$, $h_{st}$ and $h_{tr}$, it may be different from that in traditional point-to-point wireless communication systems. However, when the tag modulates bit `0', the effective channel reduces to the traditional communication channel.}

\vspace{-2mm}
\section{DoAs and Channel Gains Estimation} \label{S3}
 This section describes the procedure for the estimation of the DoAs and the channel gains. The technique is divided into the following three steps:
 i) the channel $\mathbf{h}$ in the absorptive or reflective state  is incipiently retrieved by means of LS;
 ii) by performing DFT operation on the channel $\mathbf{h}$, coarse  DoAs $[\theta_0,\theta_1]$ and gains $[h_0, \frac{h_1}{\eta}]$ can be obtained; and
 iii) with the aid of angular rotation, fine estimates of both the DoAs and the channel gains are acquired.

\textit{\textbf{{Step~1: Initial Channel Estimation}} }

 Prior to the tag modulates its own data, while the tag initially transmits $2N$ control sequences, the RF source transmits $2N$ pilots. Specifically,  the tag transmits bit `0' during the first $N$ RF symbols and bit `1' during the following $N$ RF symbols.
For $B(n)=i$,\,$i\in\{0,1\}$, we denote  $N$ RF pilots as $\mathbf{s}=[s_1, s_2, \cdots, s_N]^T$ and each element has modulus $\sqrt{P_s}$, i.e., $|\mathbf{s}_n|^2=P_s$.
Then, the received signal matrix of size $M \times N$  at the reader is
\begin{align}\label{pilot}
\mathbf{Y} &= \mathbf{h} \mathbf{s}^T+\mathbf{W},
\end{align}
where $\mathbf{W}$ is   the  $M \times N$ noise matrix.
Then, an LS estimator for the desired channel $\mathbf{h}$ is
\begin{align}\label{LS}
\mathbf{\hat{h}}^{\text{LS}}
&=\mathbf{Y} \mathbf{s} (\mathbf{s}^T \mathbf{s})^{-1} = \mathbf{h}  + \mathbf{Ws}{(N P_s)}^{-1}.
\end{align}

\textit{\textbf{{Step~2: Coarse DoAs and Gains Estimation via DFT}} }

We define the DFT matrix as
\begin{align}\label{DFTmatrix}
[\mathbf{F}]_{pq}=M^{-1}e^{-j\frac{2\pi}{M}(p-1)(q-1)},\quad  p,q \in \{1, \cdots, M\}.
\end{align}
 Then, the DFT of the channel $\mathbf{h}$ is
\begin{align}\label{DFT}
\mathbf{\hat{h}}^{\text{DFT}}=\mathbf{F}\mathbf{h},
\end{align}
whose $m$th entry can be calculated as
\begin{align}\label{DFTValue}
\mathbf{\hat{h}}^{\text{DFT}}_{m}
=&\sum_{q=1}^M \frac{1}{{M}}e^{-j\frac{2 \pi}{M}(m-1)(q-1)}\left(h_{0} e^{j \frac{2 \pi d}{\lambda}(q-1) \sin \theta_0}\right. \notag\\
&\quad \quad \quad \quad \quad \quad \quad \quad\left.+ h_1 B(n)e^{j \frac{2 \pi d}{\lambda} (q-1) \sin \theta_1}\right) \notag\\
\overset{\text{(a)}}{=}&\frac{h_{0}}{{M}}  e^{-j\frac{M-1}{2} r_0} \frac{\sin(\frac{M}{2}r_0)}{\sin(\frac{1}{2}r_0)} \notag \\
&+\frac{h_{1}B(n)}{{M}}  e^{-j\frac{M-1}{2}r_1} \frac{\sin(\frac{M}{2}r_1)}{\sin(\frac{1}{2}r_1)},
\end{align}
where (a) follows by using the formula of summation for geometric sequence, and $r_i=\frac{2 \pi (m-1)}{M}- \frac{2 \pi d}{\lambda} \sin \theta_i$ for $i\in \{0,1\}$. According to \eqref{DFTValue}, if $\frac{ Md}{\lambda} \sin \theta_i+1$ is equal to certain integer $m$,  $\mathbf{\hat{h}}^{\text{DFT}}$ has  only one non-zero item $\mathbf{\hat{h}}^{\text{DFT}}_{m}=h_i$ when %$M$ tends to infinity, i.e.,
$M \rightarrow \infty$.
 This means that the channel power is centred on only one position $m=\frac{ Md}{\lambda} \sin \theta_i+1$. Further, the DoAs and the channel gains can be separately estimated as
  \begin{equation}
 \hat{\theta}_i^{\text{DFT}}=
 \left\{
 \begin{array}{ll}
\arcsin \left(\frac{(m-1-M) \lambda}{M d} \right),&\theta_i \in [-\frac{\pi}{2},0] \\
\arcsin \left(\frac{(m-1) \lambda}{M d}\right),&\theta_i \in [0, \frac{\pi}{2}]
\end{array}
 \right. \label{CoarseDOA}
  \end{equation}
\begin{equation}
\hat{h}_i^{\text{DFT}}=\mathbf{\hat{h}}^{\text{DFT}}_{m}. \label{CoarseGain}
 \end{equation}
 However,   the actual situation is that  $\frac{ Md}{\lambda} \sin \theta_i$ is not always an integer,  where we can take $ m=\lfloor \frac{ Md}{\lambda} \sin \theta_0 \rceil +1$.

\textit{\textbf{{Step 3: Refining  Estimates Through  Angular Rotation}} }

Performing angular rotation operation yields
\begin{align}\label{rotOperation}
\mathbf{\hat{h}}^{\text{Ro}}
&=\mathbf{F} \bm{\Phi}(\Delta_i) \mathbf{h},
\end{align}
where $\bm{\Phi}(\Delta_i)=\textrm{diag}\{1,e^{j\Delta_i},\cdots,e^{j(M-1)\Delta_i}\}$ is angular rotation matrix for $\Delta_i \in [-\frac{\pi}{M},\frac{\pi}{M}]$.
Similarly, the $m$th element of $\mathbf{\hat{h}}^{\text{Ro}}$ has the form as
\begin{align}\label{est}
\mathbf{\hat{h}}^{\text{Ro}}_{m}
=&\frac{h_{0}}{\sqrt{M}}  e^{-j\frac{M-1}{2} \tilde{r}_0} \frac{\sin(\frac{M}{2}\tilde{r}_0)}{\sin(\frac{1}{2}\tilde{r}_0)} \notag \\
&+\frac{h_{1}B(n)}{\sqrt{M}}  e^{-j\frac{M-1}{2}\tilde{r}_1} \frac{\sin(\frac{M}{2}\tilde{r}_1)}{\sin(\frac{1}{2}\tilde{r}_1)},
\end{align}
where
  $\tilde{r}_i=\frac{2 \pi (m-1)}{M}- \frac{2 \pi d}{\lambda} \sin \theta_i - \Delta_i$ for $i\in \{0,1\}$.
Obviously, there always exists $\Delta_i$ which makes
\begin{align}\label{locationFine}
m= \frac{ Md}{\lambda} \sin \theta_i+\frac{M\Delta_i}{2 \pi}+1,
\end{align}
an integer.
 Then,   we can refine the corresponding parameters  as
 \begin{equation}
\hat{\theta}_i^{\text{Ro}}=
 \left\{
 \begin{array}{ll}
\hspace{-2mm}\arcsin \left(\frac{(m-1-M) \lambda}{M d} - \frac{\Delta_i \lambda}{2 \pi d}\right),& \theta_i \in [-\frac{\pi}{2},0] \\
%\\
\hspace{-2mm}\arcsin \left(\frac{(m-1) \lambda}{M d} - \frac{\Delta_i \lambda}{2 \pi d}\right),&\theta_i \in [0, \frac{\pi}{2}]
\end{array}
 \right. \label{FineDOA} \\
  \end{equation}
\begin{equation}
 \hat{h}_i^{\text{Ro}}=\mathbf{\hat{h}}^{\text{Ro}}_{m}. \label{FineGain}
  \end{equation}
\remark
The presented method is also applicable to channel estimation in multi-path or frequency-selective channels scenarios since the composite channel in the case of $B(n)=1$ can be treated as a combination of  paths $S-R$ and $S-T-R$.

\vspace{-2mm}
\section{ Cram\'{e}r-Rao Lower Bounds} \label{S4}
In this section, we compute the CRLBs for the modulus of the channel gain and the DoA estimates.
Suppose $h_0=|h_0|e^{j\omega_0}$ and $h_1=|h_1|e^{j\omega_1}$.
Let us define vector $\bm{\varphi}=[|h_0|,|h_1|, \sin \theta_0, \sin \theta_1]^T$
 and $g(\bm{\varphi})=[|h_0|,\frac{|h_1|}{\eta}, \theta_0, \theta_1]^T$. For a given $\bm{\varphi}$, the probability density function $p(\mathbf{y};\bm{\varphi})$ of  $\mathbf{y}=[\mathbf{y}^T(N+1),\cdots, \mathbf{y}^T(2N)]^T$ during $N$ consecutive RF signals  is
\begin{align}
p(\mathbf{y};\bm{\varphi})
=& \frac{\pi^{-MN} }{\sigma^{2MN}}  \prod_{n=N+1}^{2N}
e^{  -\frac{\|\mathbf{y}(n)- (\mathbf{h}_{sr}+\eta h_{st} \mathbf{h}_{tr})  s(n)\|^2 }{\sigma^{2}} }. \label{PDF}
\end{align}
 The Fisher information matrix of vector $\bm{\varphi}$ is defined as \cite{helstrom1994book}
 \begin{align} \label{FIM}
[\mathbf{I}(\bm{\varphi})]_{m,n}=-\mathbb{E} \left\{ \frac{\partial^2 \ln p(\mathbf{y};\bm{\varphi})}{\partial {\bm{\varphi}}_m  \partial {\bm{\varphi}}_n } \right \}.
 \end{align}
Let $c_0=\frac{2\pi d }{\lambda }$, $c_1=\frac{2\pi d}{\lambda} (\sin \theta_0-\sin \theta_1)$ and $c_2=\omega_0-\omega_1$,
and the Fisher information matrix and its entries  are
 \begin{align}
&\mathbf{I}(\bm{\varphi})=
\frac{2\displaystyle \sum_{n=N+1}^{2N}|s(n)|^2}{\sigma^2}
\left(
                             \begin{array}{cccc}
                               T_{11} & T_{12} & 0 & T_{14} \\
                               T_{12} & T_{22} & T_{23} & 0 \\
                               0 & T_{23} & T_{33} & T_{34} \\
                               T_{14} & 0 & T_{34} & T_{44} \\
                             \end{array}
                           \right), \label{Fisher} \\
 &T_{11}=M,\,T_{12}= \sum_{m=0}^{M-1}\cos(   c_1 m +c_2), \notag
 \end{align}
\begin{align}
 &T_{14} =c_0   |h_1| \sum_{m=0}^{M-1} m \sin( c_1m+c_2 ), \, T_{22}=M,  \notag \\
 &T_{23} = -c_0  |h_0| \sum_{m=0}^{M-1} m \sin(c_1m+c_2),  \notag \\
 &T_{33}= { c_0^2 h_0^2 (M-1)M(2M-1)}/{6},  \notag \\
 &T_{34}=c_0^2  |h_0 h_1| \sum_{m=0}^{M-1}  m^2 \cos(c_1 m+c_2), \notag \\
 &T_{44}= {c_0^2 h_1^2 (M-1)M(2M-1)}/{6}.\notag%\label{T44}
\end{align}
The proof of \eqref{Fisher} is given in Appendix.

 Afterwards, the CRLB of the estimate $\hat{g}(\bm{\varphi})$ of $g(\bm{\varphi})$ can be derived by using the equality:
 \begin{align}
 \text{CRLB}\bigg([\hat{g}(\bm{\varphi})]_{m,1}\bigg) &= \left[\frac{\partial g(\bm{\varphi})}{\partial \bm{\varphi}}\mathbf{I}^{-1}(\bm{\varphi}) \frac{\partial g(\bm{\varphi})}{\partial \bm{\varphi}}^{T}\right]_{mm} \notag \\
 &= \left[\frac{\partial g(\bm{\varphi})}{\partial \bm{\varphi}} \right]_{mm}^2 [\mathbf{I}^{-1}(\bm{\varphi})]_{mm}, \label{CRLB}
 \end{align}
 where $\frac{\partial g(\bm{\varphi})}{\partial \bm{\varphi}}= \text{diag}\left \{1,\frac{1}{\eta},\frac{1}{\cos \theta_0 },\frac{1}{\cos \theta_1 }  \right\}$.

Based on \eqref{Fisher} and \eqref{CRLB}, the CRLBs of the modulus of the channel gains and the DoAs estimates in the case of $B(n)=1$ can be respectively formulated as
\begin{equation}
\begin{split}
 &\text{CRLB}\left(|\hat{h}_0|\right) = \frac{\sigma^2 (T_{22}T_{33}T_{44}-T_{22}T^2_{34}-T^2_{23}T_{44})}{2 L_1  \sum_{n=N+1}^{2N} |s(n)|^2},  \\
 &\text{CRLB}\left(\frac{|\hat{h}_1|}{\eta}\right) =  \frac{\sigma^2(T_{11}T_{33}T_{44}-T^2_{14}T_{33}-T_{11}T^2_{34})}{2 L_1   \eta^2 \sum_{n=N+1}^{2N} |s(n)|^2},  \\
 &\text{CRLB}\left(\hat{\theta}_0\right) = \frac{\sigma^2(T_{11}T_{22}T_{44}-T_{22}T^2_{14}-T^2_{12}T_{44})}{2 L_1  \cos ^2\theta_0 \sum_{n=N+1}^{2N} |s(n)|^2} , \\
 &\text{CRLB}\left(\hat{\theta}_1\right) = \frac{\sigma^2(T_{11}T_{22}T_{33}-T_{11}T^2_{23}-T^2_{12}T_{33})}{2 L_1  \cos ^2\theta_1 \sum_{n=N+1}^{2N} |s(n)|^2},\label{CRLB13}
\end{split}
\end{equation}
where
\begin{align*}
L_1=& \bigg(T^2_{14}(T^2_{23}-T_{22}T_{33})-2T_{12}T_{14}T_{23}T_{34} \notag\\
& +(T^2_{12}-T_{11}T_{22})(T^2_{34}-T_{33}T_{44})-T_{11}T^2_{23}T_{44} \bigg).
\end{align*}
 Considering that $[\mathbf{I}^{-1}(\bm{\varphi})]_{mm}  \geq  [\mathbf{I}(\bm{\varphi})]_{mm}^{-1}$, we obtain a lower bound of CRLBs (LCRLBs) in the case of $B(n)=1$ as
  \begin{align} \label{LCRLB}
 \text{LCRLB}\bigg( [\hat{g}(\bm{\varphi})]_{m,1}\bigg)
= \left[\frac{\partial g(\bm{\varphi})}{\partial \bm{\varphi}} \right]_{mm}^2 \frac{1}{[\mathbf{I}(\bm{\varphi})]_{mm}}.
 \end{align}
Consequently, the corresponding LCRLBs can be shown as
\begin{align}
 &\text{LCRLB}\left(|\hat{h}_0|\right) =\frac{(2M)^{-1}\sigma^2}{\sum_{n=N+1}^{2N} |s(n)|^2},\label{LCRLB10} \\ &\text{LCRLB}\left(\frac{|\hat{h}_1|}{\eta}\right) = \frac{(2M\eta^2)^{-1}\sigma^2}{ \sum_{n=N+1}^{2N} |s(n)|^2}, \label{LCRLB11}  \\
 &\text{LCRLB}\left(\hat{\theta}_0\right)= \frac{3M^{-1} (2 \pi d h_0 \cos\theta_0)^{-2} \lambda^2 \sigma^2} {(M-1)(2M-1)  \sum_{n=N+1}^{2N} |s(n)|^2}, \label{LCRLB12}
\\
 &\text{LCRLB}\left(\hat{\theta}_1\right)= \frac{3 M^{-1} (2 \pi d h_1 \cos \theta_1)^{-2} \lambda^2 \sigma^2} {(M-1)(2M-1)  \sum_{n=N+1}^{2N} |s(n)|^2}. \label{LCRLB13}
\end{align}

In a similar way, the CRLBs of the modulus of the channel gain and the DoA estimates for $B(n)=0$ can be derived as \eqref{LCRLB10} and \eqref{LCRLB12} just by replacing the range of $n$ with $\{1,\cdots,N\}$, respectively.

\remark
As shown in Fig.~\ref{angle} and Fig.~\ref{gain},  the related curves of the CRLBs and LCRLBs fit well.
Accordingly, we can replace the CRLBs \eqref{CRLB13}  with  LCRLBs (\ref{LCRLB10})-(\ref{LCRLB13}), which are more straightforward for performance analysis.

\section{Simulation  Results} \label{S5}
In this section, we select $N=1$, $\sigma^2=1$ and $\eta=0.5$.
The reader is configured with $M=128$ antennas.
The DoAs are set to $\theta_0= -\frac{\pi}{4}$ and $\theta_1= \frac{\pi}{5}$.
All fading channels are modeled   as ${h}_{0},h_{st},h_{tr}  \sim \mathcal{CN}(0,1)$. Here,  the mean square error (MSE) of an estimator $\hat{x}$ is the average squared difference between the estimated result $\hat{x}$ and what is estimated $x$, i.e., $\text{MSE}(\hat{x})=\mathbb{E}\{(x-\hat{x})^2\}$.

\begin{figure}
  \centering
  \includegraphics[height=68mm,width=85mm]{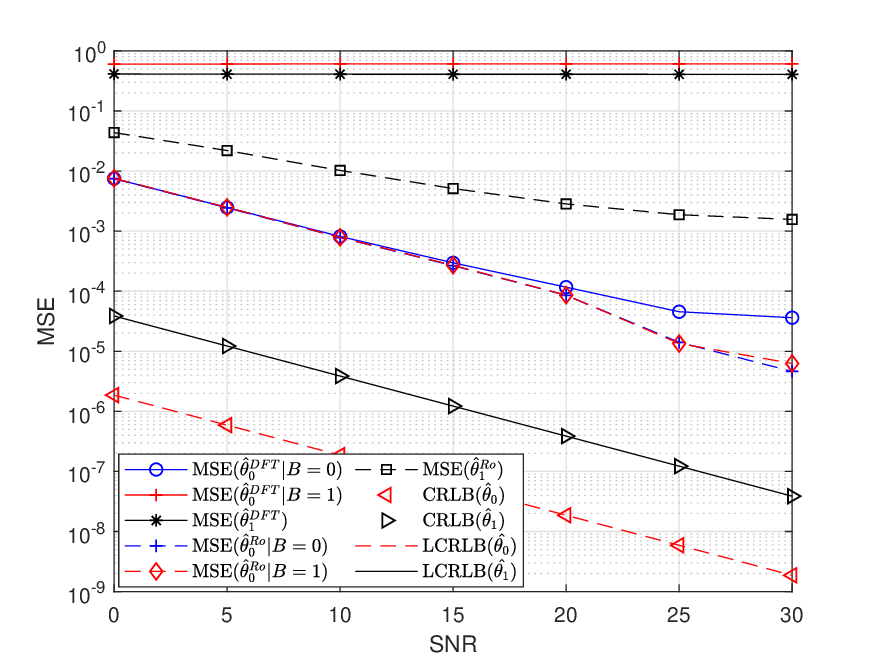}
  \caption{MSE of DoA versus transmit SNR.}\label{angle}
   \vspace{-2mm}
\end{figure}

\begin{figure}
  \centering
  \includegraphics[height=68mm,width=85mm]{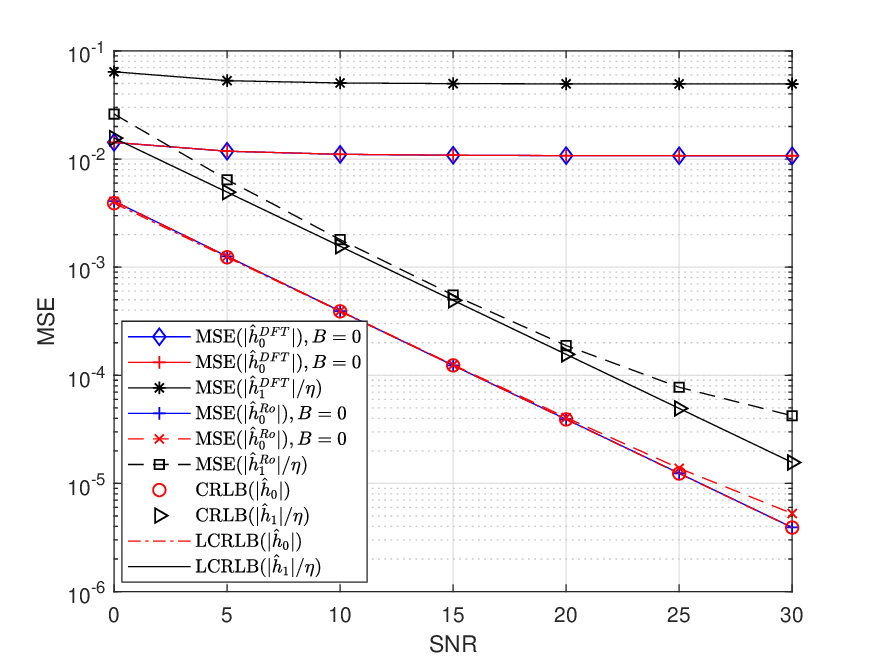}
  \caption{MSE of the modulus of channel gain versus transmit SNR.}\label{gain}
   \vspace{-2mm}
\end{figure}

Fig.~\ref{angle} displays MSEs of DoAs versus transmit SNR and  Fig.~\ref{gain} shows MSEs of the modulus of channel gains versus transmit SNR. According to our analysis in Section~\ref{S4}, the CRLBs of the estimates $\hat{\theta}_0$ and $|\hat{h}_0|$ of $\theta_0$ and $|h_0|$  in the case of $B(n)=0$ are equal to the corresponding LCRLBs in the case of $B(n)=1$. Therefore, we omit the CRLB curves in the case of $B(n)=0$.
Both MSEs and CRLBs of the DoAs and the modulus of channel gains estimates decrease with  the increase of  transmit SNR. It can be found that the corresponding CRLB curves are exactly below the corresponding MSE curves of our proposed estimator, which verifies the validity  of our theoretical derivations in  \eqref{CRLB13}.

\begin{figure}
  \centering
  \includegraphics[height=68mm,width=85mm]{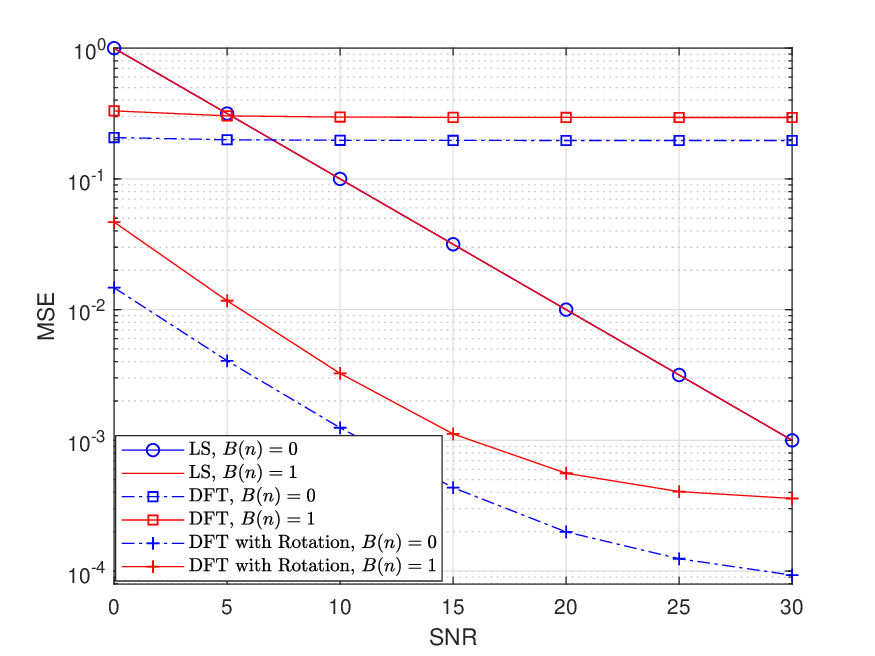}
  \caption{MSE of channel versus transmit SNR.}\label{channel}
   \vspace{-2mm}
\end{figure}

Fig.~\ref{channel} depicts MSEs of estimates of $\mathbf{h}$ versus transmit signal-to-noise ratio (SNR).
 As  expected, the DFT-based estimator with angular rotation (`+' marked curves) performs  better than that without  rotation (`$\Box$' marked curves).
We also observe that DFT-based estimator with angular rotation outperforms the traditional LS estimator (solid curves). %\new{Can you give reason why we have two flat curves?}

\begin{figure}
  \centering
  \includegraphics[height=68mm,width=85mm]{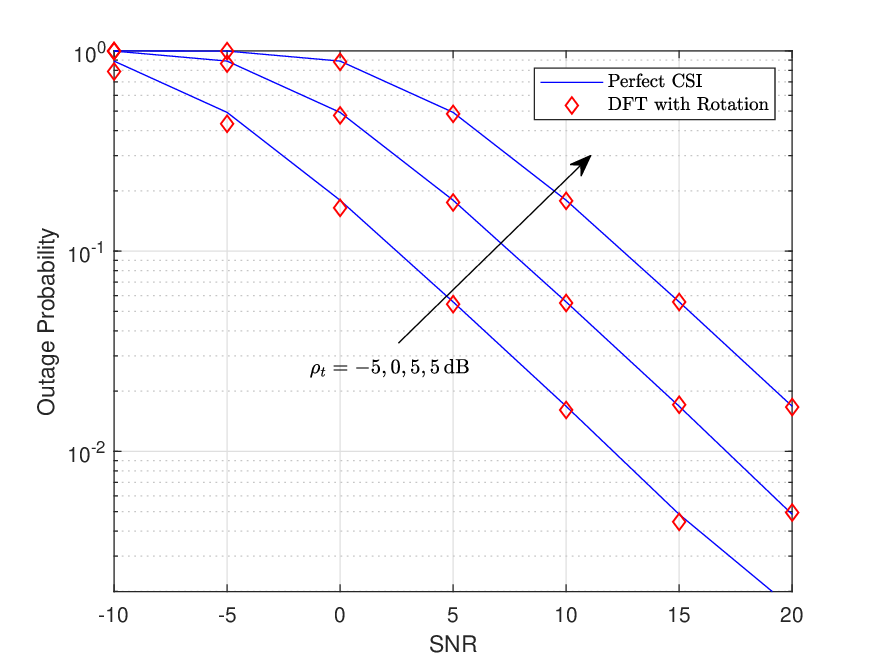}
  \caption{Outage probability versus transmit SNR.}\label{outageSC}
   \vspace{-2mm}
\end{figure}
  To show the impact of our proposed channel estimation on actual performance,  Fig.~\ref{outageSC} illustrates  outage probability versus transmit SNR when threshold $\rho_t$ is set to $-5, 0, 5\,$dB, where  we use the selection combining reception at the reader. It can be seen that the gap between outage performance with perfect CSI assumption and that with channel estimated by our proposed method is negligible over the simulated range.

\vspace{-2mm}
\section{Conclusion} \label{S6}
In this paper, we propose a channel estimation technique for an ambient backscatter communication system with a massive-antenna reader.
 We tackle the channel estimation problem
based on a signal processing standpoint as channels can be decomposed into channel gains and DoAs.
First, preliminary channel estimations at different states are acquired by the LS estimation. Coupled with angular rotation, we then obtain both the channel gains and the DoAs by conducting DFT to the estimates.   It is revealed that the proposed method separately features high estimation accuracy and low complexity compared to the LS and EVD estimators.

\vspace{-2mm}

\appendix \label{ss1}
Suppose $s(n)=|s(n)|e^{j\omega_{s_n}}$, then  taking the log-likelihood function of $p(\mathbf{y};\bm{\varphi})$ \eqref{PDF} produces
\begin{align}
&\ln p(\mathbf{y};\bm{\varphi})=-MN \ln(\pi\sigma^2)-\frac{1}{\sigma^2} \sum_{n=N+1}^{2N} \sum_{m=1}^{M} \label{PDFs} \\
  &\bigg[\big(\mathfrak{Re}\{\mathbf{y}_m(n)\}-|s(n)|\sum_{i=1}^2 \bm{\varphi}_i \cos(c_0 m \bm{\varphi}_{i+2}+\omega_i+ \omega_{s_n} ) \big)^2+  \notag \\
&\big(\mathfrak{Im}\{\mathbf{y}_m(n)\}-|s(n)|\sum_{i=1}^2 \bm{\varphi}_i \sin(c_0 m \bm{\varphi}_{i+2}+\omega_i+ \omega_{s_n})  \big)^2 \bigg],  \notag
\end{align}
where $\mathbb{E} \{\mathfrak{Re}\{\mathbf{y}_m(n)\} \}= |s(n)| \sum_{i=1}^2 \bm{\varphi}_i \cos(c_0 m \bm{\varphi}_{i+2}+\omega_i+ \omega_{s_n} ) $ and $\mathbb{E} \{\mathfrak{Im}\{\mathbf{y}_m(n)\} \}= |s(n)| \sum_{i=1}^2 \bm{\varphi}_i \sin(c_0 m \bm{\varphi}_{i+2}+\omega_i+ \omega_{s_n} ) $.

For $i\neq j, i,j\in\{1,2\}$, taking the negative second derivatives of \eqref{PDFs} yield
\begin{align} \label{Partialderivative}
 &-\frac{\partial^2 \ln p(\mathbf{y};\bm{\varphi})}{\partial {\bm{\varphi}}^2_i}=\frac{2M }{\sigma^2}\sum_{n=N+1}^{2N}|s(n)|^2, \notag\\
 &-\frac{\partial^2 \ln p(\mathbf{y};\bm{\varphi})}{\partial {\bm{\varphi}}_i \partial {\bm{\varphi}}_j}=
 \frac{2}{\sigma^2} \sum_{n=N+1}^{2N}|s(n)|^2 \sum_{m=1}^{M} \cos(c_1 m + c_2), \notag \\ %\cos[c_0 m (\bm{\varphi}_{j+2}-\bm{\varphi}_{i+2})+\omega_j-\omega_i]
 &-\frac{\partial^2 \ln p(\mathbf{y};\bm{\varphi})}{\partial {\bm{\varphi}}_i \partial {\bm{\varphi}}_{i+2}}=\sum_{n=N+1}^{2N} \frac{2|s(n)|}{\sigma^2} \sum_{m=1}^{M} \bigg(|s(n)| \bm{\varphi}_j c_0 m \times \notag \\
 &   \sin(c_1 m+c_2) +\mathfrak{Re}\{\mathbf{y}_m(n)\}
 c_0 m  \sin(c_0 m \bm{\varphi}_{i+2}  + \omega_i +\omega_{s_n}) \notag \\
 &\quad \quad \quad \quad  -\mathfrak{Im}\{\mathbf{y}_m(n)\} c_0 m  \cos(c_0 m \bm{\varphi}_{i+2}  + \omega_i +\omega_{s_n})\bigg), \notag \\
 &-\frac{\partial^2 \ln p(\mathbf{y};\bm{\varphi})}{\partial {\bm{\varphi}}_i \partial {\bm{\varphi}}_{j+2}}
 =\sum_{n=N+1}^{2N} \frac{2|s(n)|^2}{\sigma^2} \sum_{m=1}^{M} \bigg( \bm{\varphi}_j c_0 m \times \notag \\
  & \quad \quad \quad \quad \quad \quad  \sin\big(c_0 m (\bm{\varphi}_{i+2}-\bm{\varphi}_{j+2})+\omega_i-\omega_j\big)\bigg),\notag \\
  &-\frac{\partial^2 \ln p(\mathbf{y};\bm{\varphi})}{\partial {\bm{\varphi}}^2_{i+2}}=
 \sum_{n=N+1}^{2N}\frac{2|s(n)|}{\sigma^2} \sum_{m=1}^{M}\bigg(-|s(n)| \bm{\varphi}_i \bm{\varphi}_j c^2_0 m^2 \notag \\
  & \times \cos(c_1 m + c_2) + \mathfrak{Re}\{\mathbf{y}_m(n)\}
 \bm{\varphi}_i c^2_0 m^2 \cos(c_0 m \bm{\varphi}_{i+2} + \omega_i \notag \\
  &    +\omega_{s_n}) +\mathfrak{Im}\{\mathbf{y}_m(n)\}
 \bm{\varphi}_i c^2_0 m^2  \sin(c_0 m \bm{\varphi}_{i+2}  + \omega_i +\omega_{s_n})\bigg), \notag
   \end{align}
\begin{align}
  &-\frac{\partial^2 \ln p(\mathbf{y};\bm{\varphi})}{\partial {\bm{\varphi}}_{i+2} \partial {\bm{\varphi}}_{j+2}} =\bm{\varphi}_i \bm{\varphi}_j  \sum_{n=N+1}^{2N}\frac{2 c^2_0 |s(n)|^2}{\sigma^2} \times \notag \\
  &\quad \quad \quad \quad \quad \quad \quad \quad \quad \quad \quad \sum_{m=1}^{M}  m^2 \cos(c_1 m + c_2). \notag
 \end{align}
 Upon averaging the derivatives with respect to $\mathfrak{Re}\{\mathbf{y}_m(n)\}$ and $\mathfrak{Im}\{\mathbf{y}_m(n)\}$, we have entries as shown in \eqref{Fisher}.

\end{document}